\documentstyle[prl,aps,psfig,floats,twocolumn]{revtex}
\begin{document}
\draft
\twocolumn[\hsize\textwidth\columnwidth\hsize\csname
@twocolumnfalse\endcsname

\title{Invasion Percolation with Temperature and the Nature of SOC in Real 
Systems}
\author{Andrea Gabrielli$^{1,2}$, Guido Caldarelli$^{2}$ and 
Luciano Pietronero$^{2}$}
\address{$^1$ Laboratoire de Physique de la Mati\`ere Condens\'ee, 
Ecole Polytechnique, 91128-Palaiseau Cedex, France}
\address{$^2$ INFM - Unit\`a di Roma 1 "La Sapienza", P.le A. Moro 2,
00185 - Roma, Italy}

\maketitle
\date{\today}
\maketitle
\begin{abstract}
We show that the introduction of thermal noise in Invasion Percolation (IP)
brings the system outside the critical point. This result suggests a possible 
definition of SOC systems as ordinary critical systems where the critical point 
correspond to set to $0$ one of the parameters. 
We recover both IP and EDEN model, for $T \rightarrow 0$, and $T 
\rightarrow \infty$ respectively. For small $T$ we find a dynamical second 
order transition with correlation length diverging when $T \rightarrow 0$. 
\end{abstract}
]
\narrowtext
The spontaneous development of complex and fractal structures has been
studied on the basis of several models which manifest the 
property of Self-Organized Criticality (SOC)\cite{BTW}.
This concept is very intriguing and its very meaning has been highly debated.
The combination of different properties as for example stochastic and quenched 
disorder, usually brings out of the criticality. 
Henceforth, the distinction with ordinary critical phenomena 
(instead of self-organized) seems to become elusive.
In order to clarify these basic questions we consider here one
of the classical models of self-organization, the Invasion Percolation
(IP) model\cite{wilkinson}. IP describes the displacement of a fluid in a 
disordered net of random throats due to another immiscible fluid pushed with 
a vanishing pressure rate.

In this letter we study this model when a temperature-like noise $T$ 
is present. The reason for this generalization is twofold.
On one side, one is interested in the robustness of SOC with respect to 
external solicitations \cite{vergeles,CVM}. In particular we find that for 
$T\ne 0$ a finite correlation length appears and the process goes out of 
criticality.  This result (togheter with the observation that in other SOC 
model the scaling properties are limited by the ``slow driving''
\cite{zapp-vesp}) suggests a possible definition for SOC phenomena in real 
systems.  {\em A system or a dynamical process is SOC if the critical value of 
the driving parameter is $0$, instead of another real number}.
The reason why such a value makes such a large difference is because
the driving parameter of these process is always a ratio (grain of sand added
with respect to the total number of sites for the sandpiles, sites whose 
``value'' is changed for IP, DLA\cite{WS} Bak and Sneppen\cite{BS} etc.) and 
any value smaller than a certain threshold can be considered equal to $0$.
For this reason the zero value tends to occupy a much larger region of the
phase space than the other real numbers. 

On the other side IP model is the most famous and simple example of 
evolution in quenched disorder. IP dynamics of evolution
is deterministic and extremal in the sense that at each time step the 
fluid invades the environment by selecting the minimum throat available.
This limiting case is particularly instructive since the extremal dynamics is 
suitable to be studied analytically.  By using the tool of the Run Time 
Statistics (RTS) \cite{matteo,europhys}, we can approach also the more real 
stochastic case, where fluctuations affect the dynamics of invasion.

It is useful at this point to describe in some detail the IP model that 
we are studying \cite{wilkinson}: 
{\bf (a)} in a lattice of size $L$ a random number $x_i$ (extracted 
from the uniform distribution $p_0(x)=1$ for $x\in [0,1]$) is assigned 
to each bond $i$; {\bf (b)} at time $t=0$ the dynamics starts from a 
finite connected set of bonds $C_0$ 
(in the asymptotic regime the system does not depend upon $C_0$). 
$C_t$ is the connected set of bonds grown until time $t$. 
At each time step the interface $\partial C_t$ of the growing cluster 
$C_t$ is the set of non-grown bonds in contact with the cluster.
Only bonds belonging to $\partial C_t$ grow at the time step $t$;
{\bf (c)} At time $t$ the bond $i\in\partial C_t $ with the lowest 
$x_i$ grows; then $C_{t+1}=C_t \cup \{i\}$ and the interface is 
updated.  The dynamics stops when the cluster percolates the lattice.

This simple growth model develops spontaneously geometrical and dynamical 
critical features. In particular:
{\bf (1)} the asymptotic cluster is a fractal (i.e. it has an 
infinite correlation length) with fractal dimension $D_f\simeq 1.89$ in a 
$2d$ lattice, which is the same fractal dimension of the infinite cluster 
of percolation at the critical point;
{\bf (2)} the normalized histogram $\phi_t(x)$, of the  
interface variables, has the following asymptotic shape: \begin{equation}
\phi_t(x) = \frac{1}{1-p_c}\theta(x-p_c),
\label{isto}
\end{equation}
while the initial shape is obviously $\phi_0(x)=1$.
$p_c$ coincides with the critical threshold of the percolation in 
the lattice;
{\bf (3)} the asymptotic dynamics evolves for {\em critical avalanches}. 
Any bond $i$ growing at time $t$ is the {\em initiator}
of an own avalanche.
An {\em avalanche} is a temporal consecutive sequence of causally and
geometrically connected growth events starting with the growth of the 
{\em initiator}
(for a detailed definition of avalanche see e.g. \cite{BTW}). 
Note that $x$ of the initiator, due
to the shape of the asymptotic histogram and of the acceptance profile
must be $x\le p_c$ if $t$ is very large.
The size distribution $D(s;x)$ (where $x$ is the random number of the 
initiator) of the avalanche has the following behavior:
\begin{equation}
D(s;x)=s^{-\tau} f(s^{\sigma}|x-p_c|),
\label{avalanche}
\end{equation}
where $f(x)=constant$ for $x\ll 1$ and decay exponentially for $x\gg 1$ (i.e.
for $s>s_0=|x-p_c|^{-1/\sigma}$), 
with $\tau= 1.57\pm 0.03$ and $\sigma=1-\tau+2/D_f=0.49\pm 0.03$.
It may be observed that if $x=p_c$ the size distribution is a power law and 
the characteristic size diverges.
As a consequence, this kind of avalanches are called {\em critical} 
avalanches. 

We now generalize the IP model by introducing the presence of thermal noise.
The numerical study of an analogous application to the Bak-Sneppen model 
\cite{BS} can be found in \cite{vergeles} whilst the case of sandpiles has 
been considered in \cite{CVM}. 
The first effect of a finite temperature $T$ is that the deterministic 
dynamics becomes stochastic, in such a way that the larger the 
temperature the larger the stochasticity.
The definitions of $C_t$ and $\partial C_t$ in this model are the same of IP, 
but the growth rule is different: each bond $i\in\partial C_t $ has the 
following growth probability depending on the realization of the quenched 
disorder:
\begin{equation}
\eta_{i,t}(\{x\}_{\partial C_t})=\frac{e^{-\beta x_i}}{\sum_{j\in\partial C_t} 
e^{-\beta x_j}}
\label{eta}
\end{equation}
where $\beta = 1/T$ and $\{x\}_{\partial C_t}$ the realization of the quenched
disorder on the interface $\partial C_t $.
The larger is $T$ the more $\eta_{i,t}$ is 
independent on $x_i$.
Hereafter, we shall indicate with $\|C_t\|$ the number of bonds belonging to 
$C_t$, and with $\|\partial C_t\|$ the number of bonds belonging to 
$\partial C_t$.

\begin{figure}
\centerline{\psfig{file=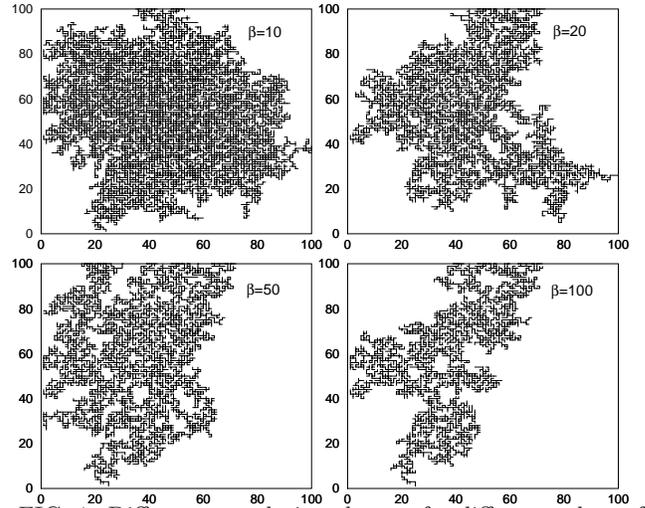,height=6.7cm,angle=-90}}
\caption{Different percolating clusters for different values of 
$\beta=\frac{1}{T}$; note that the ``fractality'' increases with $\beta$.}
\label{fig1}
\end{figure}

It is important to study the two different limits $T\rightarrow \infty$ and 
$T\rightarrow 0$.
In the first limit we have:
\begin{equation}
\lim_{T\rightarrow \infty}\eta_{i,t} = \frac{1}{\|\partial C_t\|}
\label{tinf}
\end{equation}
where $\|\partial C_t\|$ is the total number of bonds belonging to the growth 
interface at time $t$. Eq.(\ref{tinf}) means that all the bonds on the 
interface have the same probability 
to grow. This model is well known and usually called Eden model \cite{eden}: 
this dynamical growth generates a compact cluster (fractal dimension equal 
to the space dimension) with a rough surface (interface).
In the second limit we have:
\begin{equation}
\lim_{T\rightarrow 0}\eta_{i,t} = \prod_{j\in \partial C_t - \{i\}} 
\theta(x_j-x_i)
\label{t-0}
\end{equation}
where $\partial C_t - \{i\}$ means the interface $\partial C_t$ minus the 
bond $i$.
Eq.(\ref{t-0}) provides nothing else the deterministic extremal growth rule of 
IP: $\eta_{i,t}=1$ if $x_i$ is the extremal (minimum) value and zero 
otherwise.
In this paper we address mainly the study of the behavior for small 
values of $T$, i.e. the transition of the model towards IP.
In particular we will study the case of a $2d$ square bond lattice.
We started by studying some Monte-Carlo simulations of this model.
The presence of the temperature introduces a characteristic length 
$\xi(T)$ the effect of which is quite clear in Fig.\ref{fig1} where percolating 
clusters for different values of $\beta=1/T$ are shown.
The differences between the clusters can be explained 
by characterizing qualitatively the dynamical evolution of the 
growth. 

For any value of $T$, a characteristic time $t^*(T)$ exists
such that, for $t<t^*(T)$ the dynamics of the model is the IP dynamics, i.e.
even if the dynamical rule given by Eq.\ref{eta} is not deterministic, 
the effect of stochasticity 
is still negligible and the effective dynamics is almost extremal.
On the other hand for $t> t^*(T)$ the effect of the stochastic noise begins 
to be more and more important and the deviation 
from IP and then from fractality, becomes larger. 
If we suppose that $T\ll 1$, and then $t^*(T)\gg 1$, it is clear that $t^*(T)$ 
represents the correlation time of the system. 
Since one bond is removed for each time-step, $t^*(T)$ represents also the 
number of bonds $s_0(T)$ in a correlated region
of the cluster when $t\gg t^*(T)$.
This is in agreement with the idea that at $T>0$ IP is the repulsive fixed 
point of the dynamics under a spatio-temporal coarse-graining transformation, 
and the Eden model is the trivial attractive fixed point characterized 
by $T\rightarrow\infty$.
These features can be checked by looking at the dynamical evolution of the 
histogram $\phi_t(x)$. Obviously $\phi_0(x)=1$; for $t<t^*(T)$ as 
previously noted, the evolution is the same of IP, that is
$\phi_t(x)$ evolves in the step-function given by Eq.(\ref{isto}). 
At $t=t^*(T)$, $\phi_t(x)$ is a smoothened step function (the size of  
the smoothened interval around $p_c$ increases with $T$). 
For $t>t^*(T)$, because of stochasticity, 
the growth of bonds with $x$ well larger $p_c$ are permitted and 
the histogram $\phi_t(x)$ shifts towards high values of $x$.
We have measured through simulations $t^*(T)$ by measuring the time step
when $\phi_t(x)$ start to shifts and we obtain the scaling law
$t^*(T)\equiv s_0(T)\sim T^{-\gamma}$ with $\gamma=1.9\pm 0.2$.
In the following we find the same behavior theoretically and we 
link it to the correlation length of the structure. 

In order to study analytically the model, we formulate the generalization to 
stochastic growth dynamics of the Run Time Statistics (RTS) 
\cite{matteo,europhys} that we call Generalized Run Time 
Statistics (GRTS).
The usual RTS is a probabilistic technics based on the concept
of conditional probability, introduced to study IP-like dynamics, i.e.
deterministic extremal dynamics with quenched disorder.
With the GRTS approach one can solve the following problem:
suppose to fix the time-ordered path $C_t$ followed by the dynamics, and
to ignore the realization of the disorder: then 
one can compute the joint probability density function 
$P_t(\{x\}_{\partial C_t})$ of all the variables 
$x_i$ of the bonds $i$ belonging to the interface $\partial C_t$, 
conditioned to the history $C_t$.
Furthermore one can compute the conditioned probability of any possible 
next growth step. 
This joint Probability Density Function (PDF) $P_t (\{x\}_{\partial C_t})$ 
plays a central role, since from it one can compute
the probability (conditioned to the whole past history, i.e. to all 
the previous steps of the path) of any possible next growth step.
After that, one updates consequently the joint probability 
density obtaining $P_{t+1}(\{x\}_{\partial C_{t+1}})$.
Here we expose an approximated version of GRTS. 
The approximation consists in assuming that at any time-step 
the PDF can be written as 
the product of single bond density functions $p_{k,t}(x_k)$
\[P_t(\{x\}_{\partial C_t}) = \prod_{k\in \partial C_t}p_{k,t}(x_k).\]
This means that one is assuming that all the information about the history 
can be contained in the set of effective single bond density functions.
Usually this is not the case, in fact, even if we start the dynamics with
independent variables (as in this case), the information about the dynamical 
history generates correlations among the interface variables \cite{GRTS}. 
However, it can be seen that this approximation works very well even for 
IP where the the effect of this correlation is the 
maximum, due to the extremal nature of the dynamics \cite{PRE}.

Starting from the PDF's we want to compute the conditional probability
$\mu_{i,t}$ that a certain bond $i\in\partial C_t$ grows at time $t$.
Let us suppose to have the ``effective'' 
probabilities density functions 
$p_{k,t}(x_k)$ for each $k\in\partial C_t$.
The functions $p_{k,t}(x_k)$ are determined by the whole past history 
up to time $t$ (in particular, 
for $t=0$ each $p_{k,t}(x)=p_0(x)=1$ because there is no information yet 
on the dynamics).
Starting from functions $p_{k,t}(x_k)$ we write the 
conditioned probability $\mu_{i,t}$ as: 
\begin{equation}
\mu_{i,t}=\int_0^1\! ... \int_0^1 \prod_{k\partial C_t}
\left[dx_k\,p_{k,t}(x_k)\right]
\frac{e^{-\beta x_i}}{\sum_{k\in\partial C_t} e^{-\beta x_k}}
\label{mu}
\end{equation}
Eq.(\ref{mu}) provides the Growth Probability Distribution (GPD) conditioned 
to the past growth history up to time $t$. 
At this point we may iterate the procedure by updating the PDF's, that is 
by updating the ``effective'' probability density functions conditioned
to the last growth-event.
In order to do that, we have to distinguish three cases: 
(a) the last bond grown $i$, 
(b) the other bonds $j$ 
belonging to $\partial C_t$, and finally (c) the bonds just entered in the 
new interface $\partial C_{t+1}$ because of the growth of $i$:
\\
(a) in this case, $i$ does not belong to $\partial C_{t+1}$; we 
introduce the new symbol $m_{i,t+1}(x)$ analogous to
$p_{k,t}(x)$:
\begin{eqnarray}
m_{i,t+1}(x)=&&\frac{1}{\mu_{i,t}}\int_0^1\! ... \int_0^1 \prod_{k\partial C_t}
\left[dx_k\,p_{k,t}(x_k)\right]\cdot\nonumber \\
&\cdot&\frac{e^{-\beta x_i}}{\sum_{k\in\partial C_t} 
e^{-\beta x_k}}\delta(x_i-x)\;\; ;
\label{m}
\end{eqnarray}
(b) In this case we have:
\begin{eqnarray}
p_{j,t+i}(x)=&&\frac{1}{\mu_{i,t}}\int_0^1\! ... 
\int_0^1 \prod_{k\partial C_t}
\left[dx_k\,p_{k,t}(x_k)\right]\cdot\nonumber \\
&\cdot&\frac{e^{-\beta x_i}}{\sum_{k\in\partial C_t} 
e^{-\beta x_k}}\delta(x_j-x)\;\; ;
\label{p}
\end{eqnarray}
(c) Finally, we have $p_{j,t+i}(x)=p_0(x)=1$;
let us call $n_{i,t}$ the number of these bonds. 
Note that the following relations hold: 
$\|C_t\|=t$ and $\|\partial C_{t+1}\|=\|\partial C_t\|+n_{i,t}-1$. 
Here after we shall call $\Omega_t$ and $n_t$ the average values over 
different histories respectively of $\|C_t\|$ and $n_{i,t}$.

\begin{figure}
\centerline{\psfig{file=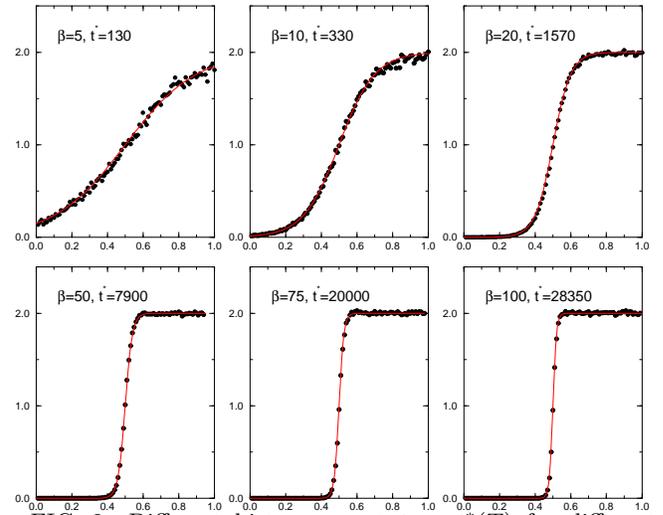,height=6.7cm,angle=-90}}
\caption{Different histograms at $t=t^*(T)$ for different 
$\beta=\frac{1}{T}$; the larger is $\beta$ the more IP-like the histogram is.
Dashed line represents Eq.(\ref{phi-as}); numerical data
are shown by filled dots.}
\label{fig2}
\end{figure}

Using Eqs.(\ref{mu}),(\ref{m}),(\ref{p}) and the rule that bonds just entered 
the interface have simply $p_0(x)=1$ as ``effective'' density function, 
one can describe from a conditional
probability point of view any possible dynamical history, knowing only 
$p_0(x)$ and the dynamical rule given by Eq.(\ref{eta}).
In \cite{europhys1,PRE2} these approach in the $T=0$ limit has been 
used to study IP to evaluate both $D_f$ and $\tau$. Now we 
use this generalized approach to study the transition towards IP 
(stochastic-extremal transition).
We introduce, then, the histogram $h_t(x)$.
$h_t(x)$ is the distribution of $x$'s on the interface at time $t$
\[\begin{array}{cc}
h_t(x)\,dx= \# \; of\; interface\; bonds\; with\; throat \in \; [x,x+dx]
\end{array}\]
If we fix a history up to time $t$, we can write:
\[h_t(x)=\sum_{i\in\partial C_t}p_{i,t}(x)\]
where the functions $p_{i,t}(x)$ must be evaluated through the ``alghorythm'' 
provided
by Eqs.\ref{mu},\ref{m},\ref{p} for the given history. Note that $\int_0^1 
dx h_t(x)\!\!=\!\!\|\partial C_t\|$.
Since the disorder is quenched (i.e. time-independent), the dynamical 
equation for $h_t(x)$ is
\begin{equation}
h_{t+1}(x)=h_t(x)-m_{i,t+1}(x)+n_{i,t}p_0(x)
\label{h-evol}
\end{equation}
It is convenient to study the normalized histogram $\phi_t(x)$ defined as
$\phi_t(x)\!\!=\!\!h_t(x)/\|\partial C_t\|$ (i.e. $\int_0^1 dx\phi_t(x)\!\!
=\!\!1$).
Since (as for IP) $\phi_t(x)$ is an almost self-averaging quantity for 
small $T$, we can 
take the average of Eq.\ref{h-evol} over all the possible histories 
in order to evaluate it.
After some algebra and some approximations, it is possible to derive the 
following equations:
\begin{equation}
\Omega_{t\!+\!1}\phi_{t\!+\!1}(x)\!=\!\Omega_t\phi_t(x)\!-\!
\Omega_t\phi_t(x)\frac{1}
{1\!\!+\!\!\Omega_{t}e^{\beta(x\!-\!1/n_t)}} \!\!+\!\!n_t
\label{phi-evol}
\end{equation}
where $\Omega_{t+1}=\Omega_t+n_t-1$.
To obtain Eq.\ref{phi-evol} we have assumed that $\Omega_t\gg 1$ 
and $e^{\beta}\gg\Omega_t$. 
Clearly the dynamical evolution of the histogram is strictly related to that of $n_t$;
in IP for $t\gg 1$ we have $n_t\simeq 1/p_c$ \cite{matteo}. 
For $t\geq t*(T)$ the evolution of $\phi_t(x)$ will be very slow (i.e. 
$|\phi_{t+1}(x)-\phi_t(x)|/\phi_t(x) 
\ll 1$), because of the increasing effect of stochasticity. 
The main effect is geometrical and it is related to the fact that, 
as the system pass from fractal to homogeneous , $n_t\rightarrow 1$ 
(it can be shown that $n_t-1$ represents the asymptotic
value of the ratio between the number of bonds belonging to the interface 
and the bonds belonging to the cluster). Here we have:
\begin{equation}
\phi_t(x)\simeq \frac{n_t}{n_t-1 + \frac{1}{\frac{1}{\Omega_t} + 
e^{\beta(x-1/n_t)}}}
\label{phi-stat}
\end{equation}
$\phi_t(x)$ is a smoothened step function centered
at $x=1/n_t$ and large $\Delta x \sim T$. 
For $t=t^*(T)\gg 1$ we have $n_t\simeq 1/p_c$, as the dynamics is IP-like
\cite{europhys1,CC}. 
Then
\begin{equation}
\phi_{t^*}(x)\simeq \frac{1}{1-p_c+\frac{p_c}{\frac{1}{\Omega_{t^*}} +
e^{\beta(x-p_c)}}}
\label{phi-as}
\end{equation}
This functions differs from eq.(\ref{isto}) only in a region of extension 
$\Delta x\sim T$ just around $x=p_c$.
The agreement between this function and the numerical data is very good 
for a wide range of $T$ (Fig.\ref{fig2}). 
From Eq.(\ref{phi-as}) and from the exponent of IP, we obtain the 
behavior of $s_0(T)$ and $\xi(T)$ for small values of $T$. 
In IP an avalanche with 
initiator different from $p_c$ of a quantity $\Delta x$ has a typical size
$s_0(\Delta x)\sim \Delta x^{-1/\sigma}$. Here we have a natural value
$\Delta x\sim T$ even for the maximal sequence of correlated growth events.
Hence $s_0(T)\sim T^{-1/\sigma} = T^{-\gamma}$ with $\gamma = 2.0\pm 0.1$ \
in agreement with the simulations.
Finally because of the fractality of IP, we have $s_0(T)\sim \xi(T)^{D}$, 
hence $\xi(T)\sim T^{-\nu}$ with $\nu=\gamma/D= 1.10\pm 0.05$.

In conclusion, we presented here a generalization of the IP model where
stochasticity, by means of a temperature-like parameter $T$ is introduced.
The model produces structures that are fractal and self-organized only
by tuning this parameter to $0$, otherwise a finite correlation length exists.
This behaviour (similar to that observed for
the Bak and Sneppen model\cite{BS} by M. Vergeles
\cite{vergeles}), supports the hypothesis that SOC models are closely related
to ordinary critical systems, where parameters have to be tuned to their
critical value.
The fundamental difference, in our opinion, is in the feasibility of this 
tuning. For SOC models, one has typically to consider limits to $0$ instead
to some other real number, the larger probability to achieve this $0$ value 
with respect to any other value is linked to the nature of the driving 
parameter that it is usually a density for such systems.

\end{document}